# Study on Applications of Supply and Demand Theory of Microeconomics and Physics Field Theory to Central Place Theory

*Benjamin Chih-Chien Nien*[*]

**Abstract:**

This paper attempts to analyze "central place theory" of spatial economics based on "supply and demand theory" in microeconomics and "field theory" in physics, and also discuss their relationship. Three most important research findings are described below. Firstly, the concept of market equilibrium could be expressed in the mathematical form of physics field theory under proper hypothesis. That is because the most important aspect of field theory model is that complex analysis is taken as a key mathematical tool. If assuming that "imaginary part" is neglected in this model, it is found that this model has the same mathematical structure as supply and demand theory of microeconomics. Secondly, the mathematical model of field theory can be applied to express clearly many concepts of central place theory, or even introduce many new concepts. Thirdly, it could also be taken as a study of combining the Hotelling Model and Moses Model for the location theory in another mathematic approach.

## 1. Introduction:

[*]Department of Economics, FuJen Catholic University, 510 Chung Cheng Rd, Hsinchuang, Taipei Hsien 24205, Taiwan, R.O.C
*email* ＜f9361001@mail.dyu.edu.tw＞

Field theory has a variety of representations, implied meanings and definitions in physics. In this paper, the field is defined as "a medium of transmitting a certain acting force in the space". The major mathematical tool is based on the concept of conformal mapping of complex analysis, for example, electrostatics, gravity field and heat transfer, etc. For this reason, this paper primarily aims to discuss the applications of field theory in spatial economics, especially for "central place theory", and to provide a firm theoretical basis for central place theory.

The application of field theory in Spatial economics was first studied in 1909 by Alfred Weber in his article: Über den Standort der Industrie (Theory of the Location of Industries), where physics isotherms, such as **isodapanes** and **isotims**[1], were introduced to analyze the positioning of firms. Since then, economist Martin Beckmann(1952) has further developed the concept of conformal mapping by analyzing the performance of transport systems and import and export trade using the Beckmann flow model. Perhaps he was also the first one who applied physics field theory into spatial economics. Moreover, Beckmann and Tönu Puu (1985) explained the concept of "Social Accounting" in Macroeconomics.

Although physics field theory is also used to analyze spatial economics in this paper, the concern of interest is central place theory developed by Christaller (1933) and Loschian (1941). The highlight of this theory is to discuss how producers are efficiently positioned in the market to realize maximum utility for both producers and consumers. In view of relatively small research efforts on the 'mathematization' of basic concepts, the major purpose of this paper is to express these basic concepts, such as "Market Range" and "Market Threshold", in the form of microeconomics and physics field theory, thereby providing a solid theoretical basis and new view for central place theory.

In addition, field theory was used by Andrew G. Pikler (1954) to explain the utility theory, especially for the concept of indifferent curve. Other than studying central place theory in another approach, this paper also analyzed Hotelling's theory from another viewpoint. Hotelling's theory, developed by Harold Hotelling (1929), was used to study

---

[1] **Isodapanes:** Contours of total transport costs used in industrial location studies. **Isotims:** Contours of transport cost for a single element in the manufacturing process.

The mathematical concept of **Consumer iso-net utility curve** in this paper is similar to **Isodapanes** and **Isotims**, but refers primarily to the net utility of consumers.

how manufacturers select location and pricing in order to obtain maximum profits, providing there are two firms and consumers are uniformly distributed within a certain market range with the same market demand. Leon Moses (1958) introduced the concept of market plane to study the selection of locations when a monopoly firm had two sources of elements. Assuming that two elements required by a manufacturer are located at two points within this plane, and the market is at another point (i.e. consumers located at that point), traditional economic concepts, such as iso-quant and iso-outlay, are used to analyze the optimal location of firms, by taking element price and transportation cost into consideration. Based on a two-dimensional market plane, this paper also used some hypothesis of Hotelling's model, namely that the consumers were uniformly distributed within the market. Hence, this paper could be considered a study that combines the Hotelling Model and Moses Model using another mathematical method.

The second part of this paper is to build upon central place theory with the concept of market supply and demand in microeconomics. It is intended to set up the relationship between market price and market distance[2] through the relationship between market price and commodity quantity. Hence, this model is named "micro-central place theory" (MCPT) in this paper. In the third part of this paper, field theory was introduced to build upon central place theory, so this model is named as "central place field theory" (CPFT) in this paper. The third part also includes three sections, of which the first section aims to prepare for central place field theory using complex analysis conformal mapping. The second section gives a formal description of mathematical models and economic meanings of "central place field theory", as well as the representation mode of market equilibrium in central place field theory. In addition, the third section discusses the mathematical relationship between central place field theory and microeconomics. It is found that when the imaginary part of central place field theory is zero, the mathematical structure is expressed by a supply and demand equation in microeconomics. In other words, because MCPT can be reflected by the relationship between commodity price and quantity in microeconomics, the demand and supply theory of microeconomics can use the field theory to express when the image part of the field is zero. The fourth part of the paper will further discuss other topics related to "central place field theory". The first section of Part 4 introduces other concepts of physics field theory into "central place field theory", and the second section discusses how to determine the prices based on the pricing of monopoly firms in CPFT. The last part describes the research potential and trend of CPF theory.

---

[2]The market distance in this paper indicates the relative distance between producers and consumers at a certain point in the market range. The market range in this paper can be considered as a maximum distance for a consumer willing to consume the product at equilibrium market price.

## 2. The Construction about Central Place Theory by Microeconomics Theory:

In 1970, Parr and Denike proposed the concept of spatial demand, whereby the relationship between commodity price and quantity in the market theory of microeconomics was replaced by the relationship between commodity quantity and market range. That is to say, commodity quantity from the Y-axis was moved to the X-axis, and market distance placed on the Y-axis. It indicated an inverse relation between the demand of consumers and the range between consumers and producers. In other words, the demand will drop with the growing range between consumers and producers due to the cost of transporting a product. This paper converts the relation between market price and quantity into a relation between market price and range, which is described below.

### 2.1 Basic assumption:

As mentioned above, central place theory was separately developed by Christaller and Loschian in 1933 and 1941. This paper adds some other assumptions for theoretical derivation, based on the following basic assumptions of central place theory by Christaller:

**(1.) Assumption on ground surface**:
   **(a)** Ground surface is a wholly flat plane. In other words, only two-dimensional analysis is theoretically possible;
   **(b)** The same degree of difficulty for traffic in all directions;
   **(c)** For homogenous transportation system on the plane, the transportation cost increases proportionally with the distance.

**(2.) Assumption of resident characteristic:**
   **(d)** All consumers are uniformly distributed on the market plane;
   **(e)** All consumers have the same demand, preference and income. Additionally, it is assumed demand $q=1$, namely, market distance $r$ can be replaced by market quantity $Q$ in Eq. (3). So, total demand quantity $Q = q_1 + q_2 + \ldots\ldots + q_n$, where n is maximum r.
   **(f)** The objective of all producers is maximum profit, and that of all consumers is maximum utility.
   It is assumed that the demand of all consumers in (e) is a single unit, as a supplement of the original assumption.

### 2.2 The Model of Microeconomic Central Place Theory (MCPT):

The relationship between MCPT model and microeconomics shall be described first of all. It is learned that general equilibrium of a market is:

$$D: P = a - b \cdot Q, \text{ where } a, b > 0 \quad (1)$$
$$S: P = c + d \cdot Q, \text{ where } c, d > 0 \quad (2)$$

According to above-specified assumptions, equations (1), (2) could be replaced by the following concept of spatial economics:

$$D: P = u - \hbar \cdot r, \text{ where } u, \hbar > 0 \quad (3)$$
$$S: P = c + d \cdot r, \text{ where } c, d > 0 \quad (4)$$

In Eq. (3), u is utility generated from one unit quantity of commodity by a consumer, $\hbar$ is unit transportation cost, r is the range between consumer and producer. According to assumptions (d) and (e), it is learned that all consumers with the same demand are uniformly distributed within a market plane. According to assumption (c), the total transportation cost for the consumer is $\hbar \cdot r$, of which h is a constant. Obviously, it increases as the distances increases. Since u and $\hbar$ are constants, only the relation of price P and range r in Eq. (3) is discussed. In other words, quantity of commodity Q can be replaced by range r.

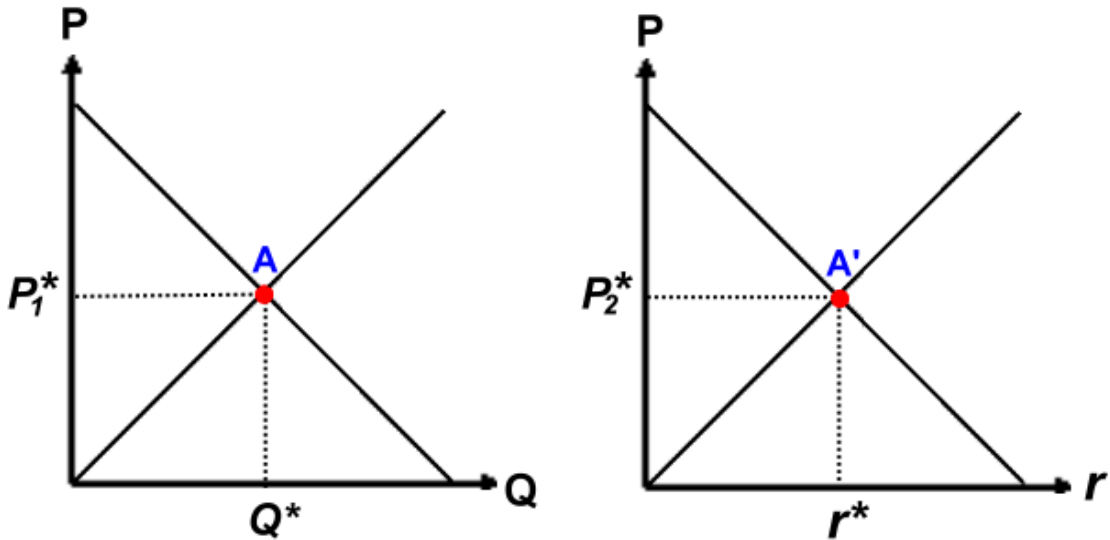

**Figure 1:** It expresses quantity of commodity Q can be replaced by range r under proper hypotheses.

Since it is assumed that every consumer consumes one-unit of a commodity, the product will increase proportionally as the market range increases. Thus, supply equation could be expressed by Eq. (4), where parameter c and d still maintain the definition of supply curve in microeconomics. That is to say, c and d in Eq. (2) and Eq. (4) have the same economic meaning. Additionally, to obtain the equilibrium solution of Eq. (3) and Eq. (4), Q in Eq. (2) could be replaced by r. Thus, the relationship between central place theory and microeconomics could be expressed by Figure 1, where the right part indicates market equilibrium range $r^*$ could be determined from the market equilibrium price $P^*$.

Net Utility *NU* of consumers and Net Revenue *NR*[3] of producers in a unit range were analyzed. Eq. (3) could be expressed by:

$$NU = (u - P^*) - \hbar \cdot r \qquad (3^*)$$

Similarly, Eq. (4) could be expressed by:

$$NR = (P^* - c) - d \cdot r \qquad (4^*)$$

where, price $P^*$ is the equilibrium price derived from Eq. (3) and Eq. (4). In other words, Eq. (3*) could be expressed as a consumer's net utility within different unit ranges, and Eq. (4*) could be expressed as a producer's net revenue within different unit ranges. Because the mathematical structures of *NU* and *NR* are the same, the following mathematical analysis is primarily dependent upon *NU*. However, under necessary situation, this paper will provide explanation on the key concepts of *NR*.

## 3. The construction about central place theory by the field theory

This section will be discussing the application of field theory in central place theory. As mentioned in the "Introduction", this model is referred to as the central place field theory (CPFT). Because the concept of conformal mapping of the complex analysis will be introduced, besides the basic assumption of the above mentioned central place theory, the first section is prepared to introduce the method of field theory. Additional assumptions and other necessary mathematical concepts will be added. In the second section, the model of CPF will be introduced and its economic meaning will be described as well. In the third section, the mathematical relationship between CPFT and microeconomics supply and demand theory will be explained to show a wide range of possible applications of "field theory" in microeconomics.

### 3.1 Complex consumer net utility potential:

**(3.)** Other Assumptions of CPFT:

**(h)** A two-dimensional market is discussed in this paper as shown in (a). So, the market distance $r$ could be expressed as $r = \sqrt{x^2 + y^2}$. It is assumed the consumer's net utility function $NU(x, y)$ could be expressed as the following component:

---

[3] Because the symbols for profits and the ratio of a circle's circumference to its diameter in economics are both $\pi$, this paper uses net revenue(*NR*) to substitute for profits($\pi$) to prevent confusion and maintain a symmetrical form.

$$Ц_x = -\frac{\partial NU}{\partial x}, \quad Ц_y = -\frac{\partial NU}{\partial y} \tag{5}$$

**(i)** It is assumed that $C$ is a simple closed contour on market plane, and $\Gamma_c$ is a contour integral of a tangent line component along $C$, $\Gamma_c$ could be expressed as:

$$\Gamma_c = \iint (\frac{\partial Ц_y}{\partial x} - \frac{\partial Ц_x}{\partial y}) dxdy = \oint_c Ц_x dx + Ц_y dy \tag{6}$$

Thus, $\Gamma_c$ is also called circulation on $C$ and it is assumed that $\Gamma_c = 0$. In other words, there isn't rotary utility force on the market plane, or the acting force generated by producer doesn't rotate.

**(j)** If assuming $\Omega_c$ is a contour integral of normal line component along $C$, $\Omega_c$ could be expressed as:

$$\Omega_c = \oint_c Ц_x dy - Ц_y dx \tag{7}$$

$\Omega_c$ is also called net flow on $C$. And, it is assumed that $\Omega_c = 0$, indicating that all utilities are provided by the producer on the market plane. If the location of producer is not included in $\Omega_c$, the relationship of utility components could be expressed by the following equation:

$$\frac{\partial Ц_x}{\partial x} + \frac{\partial Ц_y}{\partial y} = 0 \tag{8}$$

It is learned from Eq. (5) and Eq. (8) that, consumer net utility potential $NU$ is harmonic. So, Eq. (8) could be expressed as Laplace's equation below:

$$\frac{\partial^2 NU}{\partial x^2} + \frac{\partial^2 NU}{\partial y^2} = 0 \tag{8*}$$

It is also learned from Cauchy-Riemann's equation that there is surely a conjugate harmonic function $\psi(x, y)$ that leads to the analytical equation as below:

$$f(z) = NU(x, y) + i\psi(x, y) \tag{9}$$

$f(z)$ is called complex consumer net utility potential, where $z = x + iy$, $i = \sqrt{-1}$, and the modulus $|z| = \sqrt{x^2 + y^2} = r$. Thus, the following relational expression is obtained from Eq. (6) and Eq. (7):

$$\Gamma_c + i\Omega_c = \oint_c Ц_x dx + Ц_y dy + i\oint_c Ц_x dy - Ц_y dx = \oint_c (Ц_x - iЦ_y)(dx + idy) \tag{10}$$

Where, $\Gamma_c + i\Omega_c = \oint_c f'(z)dz$. In view of z plane discussed hereto, the following equation is obtained from Eq. (5):

$$\hbar = -\nabla NU = -\frac{\partial NU}{\partial x} + i\frac{\partial \psi}{\partial x} = -\overline{\frac{df}{dz}} = -\overline{f'(z)} \tag{11}$$

Where $\hbar$ is called market flux, but also a consumer's unit transportation cost as mentioned above. The meanings will be discussed in Section 4.1.

In addition, its scale is:

$$Ш = |\hbar| = \sqrt{Ш_x^{\,2} + Ш_y^{\,2}} = \left|-\overline{f'(z)}\right| = |f'(z)| \qquad (12)$$

**3.2 The model of general central place field theory (CPFT):**

Section 3.1 discussed mathematical issues required to set up CPFT, namely mathematical preparation for inducing conformal mapping into CPFT. This section is intended to discuss the economic meaning of CPFT. According to assumptions (a) and (b), a utility field will be set up if a producer is located within a region densely occupied by consumers. Firstly, the concept of a demand cone[4] is used to express the relationship between MCPT and CPFT. As shown in Figure 2, a demand cone is formed if the demand curve rotates 360° along the producer origin.

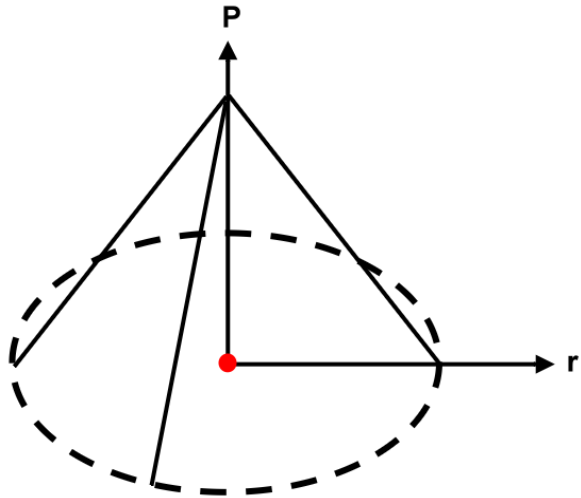

**Figure 2:** It indicates a demand cone is formed if the demand curve rotates 360 degree along the producer (origin), and the area formed on the market panel is called "utility field".

Correspondingly, the concept of a supply cone is also deduced in this paper. A revenue field may be set up by a producer within a region densely occupied by consumers. The following paragraphs will describe how these concepts are expressed by conformal mapping.

---
[4] "Demand cone", developed by Parr and Denike in 1970, was originally used to express the relationship between commodity quantity and market range in central place theory. In this paper, market quantity is replaced by market price.

Let $z = r \cdot e^{i\theta}$. To obtain Eq. (9), $f(z)$ is rewritten into $f(z) = k \cdot \ln z$, where k is a random constant. If assuming it is rewritten into Eq. (13), the following equations could be established:

$$f(z) = (u - P^*) - \hbar \cdot \ln(r \cdot e^{i\theta}) = (u - P^*) - \hbar \cdot \ln r + i\hbar\theta \quad (13)$$

$$NU(x, y) = (u - P^*) - \hbar \cdot \ln r \quad (14^*)$$

$$\psi(x, y) = \hbar \cdot Argz, \quad where \quad Argz = \theta \quad and \quad 0 \leq \theta \leq 360° \quad (15)$$

Corresponding to the concept of consumer net utility function in Eq. (14*), the mathematical expression of producer net revenue function is expressed below:

$$NR(x, y) = (P^* - c) - d \cdot \ln r \quad (16^*)$$

From demand equation and supply equation in Eq. (3),(4), it is learnt that Eq. (14*) and Eq. (16*) can also be rewritten into Eq. (14) and Eq. (16), where $P$ and $r$ are variables. Eq. (14) and Eq. (16) are called the demand equation and the supply equation of CPF theory. Thus, $P^*$ is the equilibrium price derived from Eq. (14) and Eq. (16).

$$P = u - \hbar \cdot \ln r \quad (14)$$

$$P = c + d \cdot \ln r \quad (16)$$

The important terms in the aforementioned equations imply key economic meanings, which are mainly derived from field theory and central place theory:

**Market potential range curve[5]:** It is defined as a curve obtained from Eq. (14). If $P = 0$, the maximum radius of circle $r^*$ is called the market potential range curve. As illustrated in Figure 3, it is a maximum circle on a market plane representing the potential range of consumers within the market. Thus, if $P = 0$ is substituted into Eq. (14), and may derive market potential range $\hat{r} = \exp(\frac{u}{\hbar})$.

**Consumer iso-net utility curve:** A concentric curve with an origin as the center of circle. It indicates that consumer's utility along the curve is the same. As shown in Figure 3, the utility nearer the center is higher. The mathematical expression is shown in Eq. (14*). The market potential range curve is a consumer iso-net utility curve of longest diameter.

**Producer iso-net revenue curve:** A concentric curve with an origin as the center of circle. It indicates that producer's profits along the curve are the same, and become higher if nearer the center. The mathematical expression is shown in Eq. (16*).

**Market flux line[6]:** A line used to express the acting force and direction of a producer. It

---

[5] As shown in Figure 2, demand curve has a negative slope. So, if price is zero, it will meet r-axis to obtain a market potential range curve. However, no such a concept of supply curve can be obtained since it has a positive slope.

[6] The market flux line crosses orthogonally to Consumer iso-net utility curve, while another form of market

crosses the Consumer iso-net utility curve and Producer iso-net revenue curve

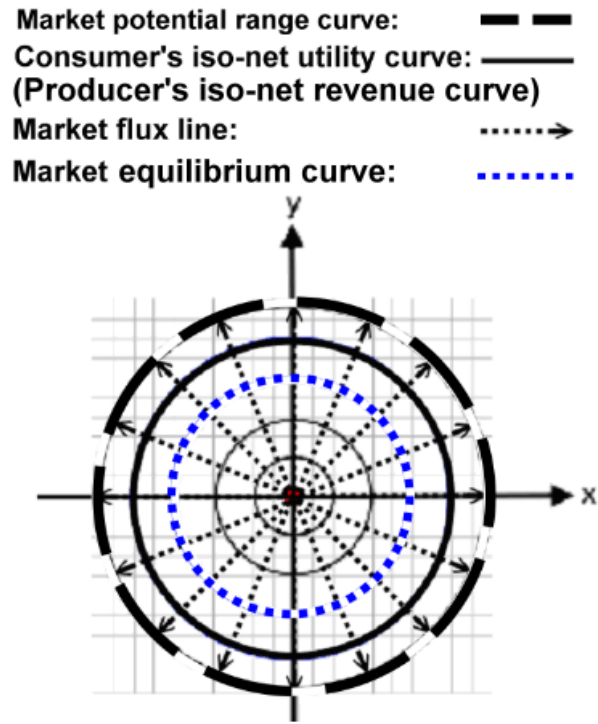

**Figure 3:** It presents some basic concepts about utility field and other related ideas.

**Market range curve:** It's also called market equilibrium curve. It represents equilibrium market price derived from Eq. (14) and (16). It can be considered as a maximum range for a consumer willing to consume the product at equilibrium market price, but also a maximum output range for a producer willing to produce the product at equilibrium market price. As expressed in Eq. (14*) or (16*), it is a curve along $NU = 0$ (or $NR = 0$). The equilibrium market range and market price is:

$$r^* = \exp\left(\frac{u-c}{\hbar+d}\right) \quad (17)$$

$$P^* = u - \hbar \cdot \left(\frac{u-c}{\hbar+d}\right) \quad (18)$$

It is also shown in Figure 4 from the angle of demand cone and supply cone. Disk B is called a disk with point B as the center of disk. Since disk B is a circle formed by market equilibrium range $r^*$, and surrounds the circumference, hence, it presents $NU = 0$ or $NR = 0$.

---

flux line crosses orthogonally to Producer iso-net revenue curve, with the mathematical expression as $\tilde{\psi}(x, y) = d \cdot Argz$. Since no any difference is found from the market flux of "supply side" and demand side, only market flux of demand side is described in this paper.

**Consumer surplus cone:** As compared with the concept of consumer surplus in microeconomics, a cone with point c as apex is called a consumer surplus cone, if disk B is the base, as shown in Figure 4. The volume is expressed by $V_{c.s}$. According to the formula of cone volume, its volume is：

$$V_{c.s} = \frac{1}{3} \cdot \pi \cdot (r^*)^2 \cdot (u - P^*) \tag{19}$$

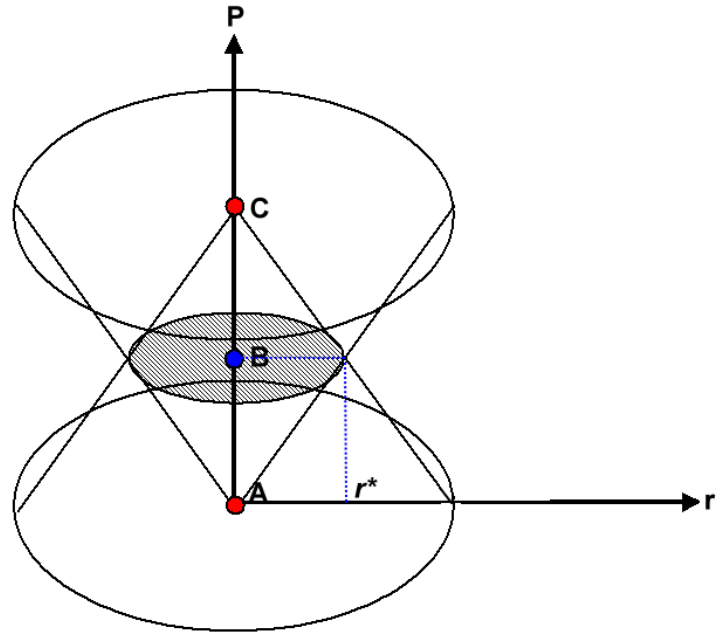

**Figure 4:** It explains the relationship between consumer surplus and producer surplus by demand cone and supply cone. Disk B is called a disk with point B as the center of disk and its circumference is called market range curve.

**Producer surplus cone:** As compared with the concept of producer surplus, a cone with point A as apex is called a producer surplus cone, if disk B is the base. The volume is expressed by $V_{p.s}$:

$$V_{p.s} = \frac{1}{3} \cdot \pi \cdot (r^*)^2 \cdot P^* \tag{20}$$

**Market threshold curve:** A curve used to express minimum market range demand as required by the producer to maintain the production activity. This concept will be expressed by microeconomics. As shown in Figure 5, if assuming a producer's average cost declines with the growing scope of service, namely, it reduces as production quantity grows, the demand curve will cross AC curve at points A and B. Point A is called a market threshold point. Some loss may occur if producer's price exceeds $P_1$, or if producer's supply exceeds $r_2$ because the average cost is highest than average profit.

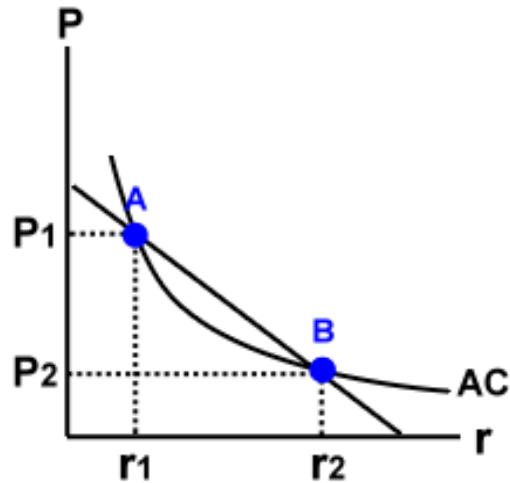

**Figure 5:** Assuming a producer's average cost declines with the growing scope of service, then it will cross AC curve at points A and B. Point A is called a market threshold point.

Also, in the CPFT model, according to assumption (g), the market threshold curve is equal to the market range curve in CPFT.

**3.3 The mathematical relationship between microeconomic and CPFT:**

It is assumed an imaginary part (let $y=0$) is neglected in Eq. (3*), namely, $NU(x)$ relates only to x. According to Eq. (8*), $NU''(x) = 0$, after integrate two times this function, we can get $NU(x) = ax + b$, where $a$ and $b$ are random constants. According to Eq. (3*), $NU(r) = (u - P^*) - \hbar \cdot r$. As modulus $|z| = r = \sqrt{x^2 + y^2}$ and $y = 0$, the following equation is obtained:

$$NU(x) = (u - P^*) - \hbar \cdot x \qquad (3^{**})$$

Where, $NU(x)$ is called the consumer net utility point[7]. According to the relation of Eq. (3*) and (4*), there is also a producer net revenue point:

$$NR = (P^* - c) - d \cdot x \qquad (4^{**})$$

As shown in Figure 6, if the left-hand of Figure 6 is overlooked along P-axis, the demand curve and supply curve will be overlapped into a line with the x-axis, i.e. right-hand figure, and point A is the intersection point of the demand curve and the supply curve, i.e. market equilibrium point.

---

[7] "Consumer net utility point" –this concept may correspond to Consumer iso-net utility curve as described in Section 3.2. As shown in right-hand of Figure 6, broken line on x-axis can be considered as a market flux line. In other words, if imaginary part isn't zero, market flux line in Figure 6 will rotate 360° along the origin, such that Consumer net utility point turns into Consumer iso-net utility curve as illustrated in Figure 2.

According to Eq. (3*) and (4*) in Section 2.2, market distance r could be replaced by market quantity Q, namely, Eq. (1) and (2) can be rewritten into:

$$NU = (a - P^*) - b \cdot Q \qquad (1^*)$$
$$NR = (P^* - c) - d \cdot Q \qquad (2^*)$$

Thus, the mathematical relationship between CPFT and microeconomics is clearly shown.

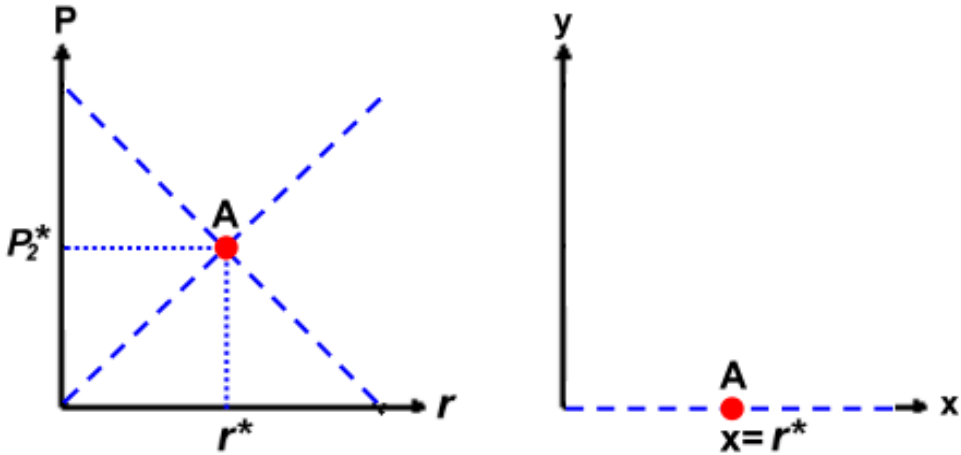

**Figure 6:** If the left-hand of Figure 6 is overlooked along P-axis, the demand curve and supply curve will be overlapped into a line with the x-axis, i.e. right-hand figure, and point A is the intersection point of the demand curve and the supply curve, i.e. market equilibrium point.

## 4. Further discussion of the Central Place Field Theory:

This part will further discuss CPFT, for example, how to deduce other concepts of field theory into CPF theory, and how to determine price in the CPF model using the pricing mode of monopoly firms. With a view to commonly used pricing methods of monopoly market manufacturers--"MR (marginal revenue) = MC (marginal cost)"--for commodity price and quantity, this paper also applied this method into the CPF model.

### 4.1 Discussion of some other physics concepts applied in CPFT:

Along with heat transfer or electrostatics, other concepts could also be inducted into CPFT model:

According to Eq. (11), market flux $\hbar$ is also expressed as:

$$\hbar = -\frac{dNU}{dr} \qquad (11^*)$$

This equation will be used to discuss the concepts of market flow and market resistance:

**Market flow:** It is defined as:

$$H = A \cdot \hbar = -(2\pi \cdot r) \cdot \frac{dNU}{dr} \tag{21}$$

Where, H is market flow, and A is the market range of market field. As for CPFT, $A = 2\pi \cdot r$; boundary value condition in Eq. (21) is $NU = NU_0$ when $r = r_0$; $NU = NU_i$ when $r = r_i$; the solution of Eq. (21) is:

$$H = \frac{2\pi \cdot (NU_0 - NU_i)}{\ln(r_i / r_0)}, \text{ where } r_0 < r_i \tag{21*}$$

The concept of market resistance is derived from Eq. (21*):

**Market resistance:** *R* represents market resistance, showing the degree of resistance for market flux. The mathematical expression is as follows:

$$R = \frac{\ln(r_i / r_0)}{2\pi} \tag{22}$$

According to Eq. (21*), the relationship between market resistance and market flow is expressed below:

$$R = \frac{NU_0 - NU_i}{H} = \frac{NU_0 - NU_i}{A \cdot \hbar} \tag{22*}$$

**4.2 Discussion of the firm make price by Monopoly Market Method:**

This section will discuss how to determine the price by using the pricing method in the microeconomics monopoly market in the CPFT model. While monopoly firms often determine the commodity quantity in the case of *MR=MC*, and then obtain a market price by substituting it into an average revenue equation, this paper strives to analyze how to apply this pricing method to CPFT.

As for producers, it is assumed the average revenue equation is the demand equation (14); and the average cost equation is the supply equation (16).

The total revenue equation is Eq. (14) multiplied by $\pi \cdot r^2$:

$$P \cdot \pi \cdot r^2 = u \cdot \pi \cdot r^2 - \hbar \cdot \pi \cdot r^2 \ln r \tag{23}$$

Then, per unit, the circle of marginal revenue is:

$$\frac{dP \cdot \pi \cdot r^2}{dr} = 2u \cdot \pi \cdot r - 2\hbar \cdot \pi \cdot r \cdot \ln r - \hbar \cdot \pi \cdot r \tag{24}$$

If removing $2\pi \cdot r$, the marginal revenue is:

$$P = u - \hbar \cdot \ln r - \frac{\hbar}{2} \tag{25}$$

Similarly, the marginal cost can be obtained from Eq. (16):

$$P = c + d \cdot \ln r + \frac{d}{2} \quad (26)$$

If the marginal cost equation (26) is substituted into Eq. (25), i.e. $MR = MC$, the equilibrium market range is:

$$r_m = \exp\left(\frac{u-c}{\hbar+d} - \frac{1}{2}\right) \quad (27)$$

If $r_m$ is substituted into Eq. (14), the equilibrium market price is:

$$p_m = u - \hbar \cdot \left(\frac{u-c}{\hbar+d} - \frac{1}{2}\right) \quad (28)$$

If compared with equilibrium price and range by substituting "supply equation (16) into demand equation(14)", the equilibrium market range is shown in Eq. (17): $r^* = \exp\left(\frac{u-c}{d+\hbar}\right)$, and the equilibrium market price in Eq. (18): $p^* = u - \hbar \cdot \left(\frac{u-c}{d+\hbar}\right)$.

## 5. Conclusion:

This part provides an overall view of the contents of this paper, and discusses the direction of research on the topic. While central place theory has always focused on how to arrange different producers in a limited market range to make both consumers and producers realize maximum utility, comparatively, there has been little research concerning mathematical and theoretical representation of some basic concepts of central place theory, such as "Market Range" and "Market Threshold", etc. The most important aspect of research has been to introduce physics field concepts into the aforementioned concepts. More importantly, it uses another method, "field", to express the concept of market equilibrium in microeconomics.

In addition, the relationship between field theory models and microeconomics supply and demand theory is also disclosed. When the mathematical model of the field theory only discusses real part (when the imaginary part is zero), this model is a microeconomics market supply function (market demand function). Thus, this paper may not only has positive impact upon spatial economics, but also provides a new vision of existing microeconomics. While field theory is introduced into central place theory to explain basic concepts and mathematical expressions, in this paper, other concepts of physics field theory are inducted and incorporated into central place theory to expand this theory's substantial content.

As this paper has discussed only the situation involving a single manufacturer, it is worthy to discuss the market situation with 2 or even n producers, and also to consider how to combine this theory with the existing central place theory for efficient arrangement in a market range.

Besides, CPFT shall also be taken back to the field of economics. For example, Andrew G. Pikler once applied field theory into utility theory of microeconomics in 1954, namely, analysis of indifferent curves. However, his works only analogized the concept of physics "iso-potential curve" into indifferent curves, thus, further discussion on the economic meanings and development on the mathematical expressions are worth researching. Also, this paper could be considered as a study combining Hotelling model and Moses model. In the future, more research efforts shall also be made to combine the ideas in this paper into Hotelling's theory, and discuss how to apply CPF theory into international trade.

**Acknowledgements:**

I am very grateful to Prof. Chen, Nen-Jing and Wu, Cheng-Tai for suggestions that greatly improved the structure of the paper.